\documentstyle[sprocl]{article}

\input{psfig}

\def\be{\begin{equation}}
\def\ee{\end{equation}}
\def\bea{\begin{eqnarray}}
\def\eea{\end{eqnarray}}

\begin{document}
\input epsf
\input{psfig}

\title{A Brief Review of Three-Family Grand Unification\\
               in String Theory
 \footnote{Report-no: HUTP-98/A059, NUB 3184}
 \footnote{Talk presented at PASCOS'98}
 \footnote{This is a short version of the review article \cite{review}.}}

\author{Zurab Kakushadze}

\address{Lyman Laboratory of Physics, Harvard University, Cambridge, MA 02138\\
  and\\
Department of Physics, Northeastern University, Boston, MA 02115\\
Email: zurab@string.harvard.edu}

\maketitle\abstracts{We briefly review the status of three-family grand unified string models.}

{}If string theory is relevant to nature, it must possess a vacuum
whose low energy effective field theory describes the Standard Model of
strong and electroweak interactions. The question whether such a string vacuum exists
is difficult to answer as the space of classical string vacua has a very large degeneracy, and there lack objective criteria that would select a particular string vacuum among the numerous possibilities. One might expect non-perturbative string dynamics to lift, partially or completely, this degeneracy in the moduli space. If this lifting is complete, then a thorough
understanding of string dynamics may be sufficient to find a complete
description of our universe. {\em A priori}, however, it is reasonable to suspect that non-perturbative dynamics may {\em not} select the unique vacuum, but rather pick out 
a number of consistent vacua, some in four dimensions with completely broken supersymmetry. 
Then our universe would be realized as one of the consistent vacua in this (probably large) set. 
If so, then we would need to impose some additional, namely, 
phenomenological constraints to find the string vacuum where we live. 
This approach has been known as ``superstring phenomenology''. The latter
 must still be augmented with understanding of non-perturbative dynamics 
as, for instance, superstring is believed not to break supersymmetry perturbatively.   

{}It might seem, at least naively, that there is more than enough 
phenomenological data to identify the superstring vacuum corresponding to 
our universe. It is, however, not known how to fully embed the Standard 
Model into string theory with all of its complexity, so one is bound to 
try to incorporate only a few phenomenologically desirable features at a 
time (such as, say, the gauge group, number of families, {\em etc.}). Since 
such constraints are not extremely stringent, 
this ultimately leads to numerous possibilities for embedding the Standard 
Model in superstring that {\em a priori} seem reasonable \cite{rev}. Thus, 
to make progress in superstring phenomenology, one needs as restrictive 
phenomenological constraints as possible.  It might be advantageous to impose
such phenomenological constraints taking into account 
specifics of a particular framework for superstring model building.   

{} In the past decade, the main arena for model building within the context
of superstring phenomenology has been perturbative heterotic superstring. 
The reason is that such model building is greatly facilitated by existence
 of relatively simple rules (such as free-fermionic \cite{FFC} and orbifold
\cite{Orb,NonAbe} constructions). Moreover, many calculational tools 
(such as, say, scattering amplitudes and rules for computing 
superpotentials \cite{scatt})
are either readily available, or can be developed for certain cases of 
interest. However, recent developments in understanding four dimensional Type I vacua \cite{typeI} have opened up a possibility for their phenomenological exploration
as well. In particular, some steps toward phenomenologically oriented model building within perturbative Type I compactifications have already been made \cite{typeIa}. Some of these four dimensional Type I vacua are non-perturbative from the heterotic viewpoint.

{}To be specific, let us concentrate on perturbative heterotic superstring. Within this framework the total rank of the gauge group (for $N=1$ space-time supersymmetric models) is 22 or less. After accommodating the Standard Model of strong and electroweak interactions (with gauge group $SU(3)_c \otimes SU(2)_w \otimes U(1)_Y$ whose rank is 4), the left-over rank for the hidden and/or horizontal gauge symmetry is 18 or less. The possible choices here are myriad \cite{standard} and largely unexplored. The situation is similar for embedding unification within a {\em semi-simple} \cite{semi} gauge group $G\supset SU(3)_c \otimes SU(2)_w \otimes U(1)_Y$ ({\em e.g.}, $SU(5)\otimes U(1)$).

{}The state of affairs is quite different if one tries to embed grand unification within a {\em simple} gauge group  $G\supset SU(3)_c \otimes SU(2)_w \otimes U(1)_Y$. Thus, an adjoint 
(or some other appropriate higher dimensional) Higgs field must be present 
among the light degrees of freedom in effective field theory to break the 
grand unified gauge group $G$ down to that of the Standard Model. 
In perturbative heterotic superstring such states in the massless 
spectrum are compatible with $N=1$ supersymmetry and chiral fermions 
only if the grand unified gauge group is realized via a current algebra
 at level $k>1$ (see, {\em e.g.}, Ref \cite{Lew}). This ultimately leads
 to reduction of total rank of the gauge group, and, therefore, to 
smaller room for hidden/horizontal symmetry, which greatly limits 
the number of possible models. 

{}The limited number of possibilities is not the only distinguishing feature 
of  grand unified models in superstring theory. Grand unified theories (GUTs) 
possess a number of  
properties not shared by superstring models with either the Standard Model or 
a semi-simple gauge group. One of such properties concerns the gauge coupling
 unification problem in superstring theory \cite{dienu}. Thus, the strong 
and electroweak couplings 
$\alpha_3$, $\alpha_2$ and $\alpha_1$ of 
$SU(3)_c\otimes SU(2)_w\otimes U(1)_Y$ in 
the minimal supersymmetric Standard Model (MSSM) unify at the GUT scale 
$M_{GUT}\sim 10^{16}~{\mbox{GeV}}$ \cite{coupling} at the value of 
$\alpha_{GUT}\sim 1/24$. 
Running of these couplings is schematically shown in Fig.1a as a function of 
the energy scale $E$. For comparative purposes, a dimensionless gravitational 
coupling $\alpha_{G} =G_N E^2$ is introduced, where $G_N$ is the Newton's 
constant.
In string theory 
the unification demands that all couplings meet at a single scale. 
Note that the gravitational coupling becomes equal $\alpha_{GUT}$ at a scale
roughly two orders of magnitude higher than $M_{GUT}$ (Fig.1a).
Several possible approaches have been proposed to reconcile this apparent 
discrepancy \cite{dienu}, some of which are listed below:\\ 
$\bullet$ The subgroups of $SU(3)_c\otimes SU(2)_w\otimes U(1)_Y$ unify 
into a single GUT gauge group $G$, and the gauge coupling of $G$ meets 
with $\alpha_G$ as shown in Fig.1b.\\ 
$\bullet$ The gauge group remains that of the MSSM all the way up, but  
some extra (compared with the MSSM spectrum) matter multiplets 
are present which 
change the
running of the couplings so that their unification scale is pushed up to meet
$\alpha_G$. Such a  scenario requires a judicious choice of the extra fields, 
as well as their masses \cite{dienf}. The situation is similar for the 
models with 
other semi-simple gauge groups.

\begin{figure}[t]
\centerline{\psfig{figure=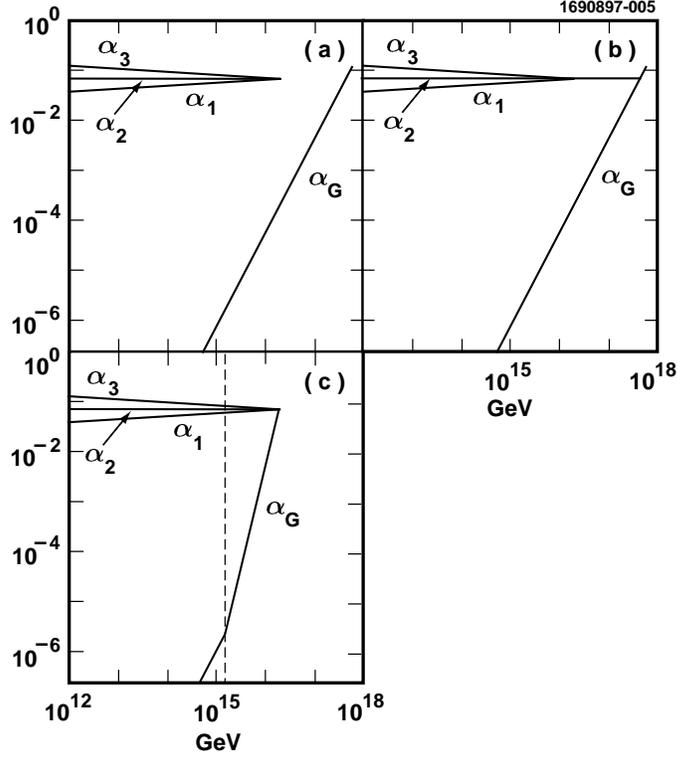,height=4in}}
\caption{Running of the MSSM couplings vs. dimensionless gravitational 
coupling. }
\end{figure}

\noindent
$\bullet$ Another possibility is suggested by $M$-theory \cite{witten}, 
where 
the graviton can propagate in a 5-dimensional space-time (the bulk), while 
the gauge and 
matter fields live on 4-dimensional ``walls''. Below the energy scale of 
the order of the inverse
``thickness'' of the bulk, all fields propagate in 4 space-time 
dimensions \cite{witten}. Above this scale, however, the gauge and matter
fields propagate in 4 dimensions, while the graviton propagates in
5 dimensions. As a result, $\alpha_G$ runs faster and catches up with the
strong and electroweak couplings as shown in Fig.1c.\\
$\bullet$ More recently the observation of  \cite{witten} has been modified into a new 
proposal \cite{TeV} where the gauge and gravitational couplings can unify at the 
string scale as low as 1 TeV. In this proposal the gauge theory lives in the world-volume of D-branes \cite{Pol} embedded in a 10 or 11 dimensional space-time. Gravity, however, propagates in the bulk of the space-time. Just as in the previous case, above certain energies 
$\alpha_G$ runs faster than $E^2$, and can get strong around 1 TeV.

{}The simplest way to obtain higher-level current algebras (required for 
GUT embeddings) in perturbative heterotic superstring is via the 
following construction. Start from a 
$k$-fold product $G^{\otimes k}=G\otimes G\otimes \cdots \otimes G$ of the grand 
unified gauge group $G$ (of rank $r$) realized via a level-1 
current algebra. The diagonal subgroup 
$G_{diag}\subset  G^{\otimes k}$ is then 
realized via a level $k$ current algebra. 
Note that the rank of the gauge group is reduced 
from $kr$ to $r$. As far as the 
Hilbert space is concerned, here we are identifying the states
under the ${\bf Z}_k$ cyclic symmetry of the $k$-fold product  
$G^{\otimes k}$. This is nothing but 
${\bf Z}_k$ orbifold action, namely, modding out by the outer automorphism.

{}An immediate implication of the above construction is a rather 
limited number of possibilities. For example, for a grand unified gauge group 
$G=SO(10)$ with, say, $k=3$, the left-over rank (for the hidden and/or 
horizontal gauge symmetry) is at most 7 ($=22-3\times 5$). This is to be 
compared with the left-over rank 18 in the case of the Standard Model 
embedding. Taking into account that the number of models grows (roughly)
 as a factorial of the left-over rank, it becomes clear that grand 
unified model building is much more restricted than other embeddings.

{}Since the desired adjoint (or higher dimensional) Higgs fields are allowed 
already at level $k=2$, multiple attempts have been made in the past 
several years to construct level-2 grand unified string models. 
None of them, however, have yielded 3-family models. The first $SO(10)$ 
string GUT realized via a level-2 current algebra was obtained by 
Lewellen \cite{Lew} within the framework of free-fermionic construction 
\cite{FFC}. Soon after Schwartz \cite{Sch} generalized Lewellen's 
construction and obtained an $SU(5)$ level-2 string GUT 
within the same framework. Both of these models have four chiral 
families. Multiple attempts have been made ever since to construct
 three-family string 
GUTs realized via level-two current algebras within free-fermionic
 construction \cite{CCHL} and within the framework of 
symmetric \cite{AFIU} as well as asymmetric \cite{Erler}
Abelian orbifolds \cite{Orb}. Finally, three of us tried 
non-Abelian orbifolds \cite{kst0} within 
both free-fermionic and bosonic formulations \cite{NonAbe}. 
There is no formal proof that 3-family models cannot be 
obtained from level-2 constructions, but one can intuitively 
understand why attempts to find such models have failed. In the 
$k=2$ construction the orbifold group is ${\bf Z}_2$. The numbers 
of fixed points in the twisted sectors, which are related to 
the number of chiral families, are always even in this case. 
This argument is not meant 
to be rigorous, but to illustrate the matter.     

{}Thus, it is natural to consider $k=3$ models. The orbifold action in 
this case is ${\bf Z}_3$, and one might hope to obtain models with 3 
families as the numbers of fixed points in the twisted sectors are 
some powers of 3. The level-3 model building appears to be more involved 
than that for level-2 constructions. The latter are facilitated by 
existence of the $E_8\otimes E_8$ heterotic superstring in 10 
dimensions which explicitly possesses ${\bf Z}_2$ outer automorphism 
symmetry of the two $E_8$'s. Constructing a level-2 model then can 
be carried out in two steps: first one embeds the grand unified gauge 
group $G$ in each of the $E_8$'s, and then performs the outer 
automorphism ${\bf Z}_2$ twist. In contrast to the $k=2$ 
construction, $k=3$ model building requires explicitly 
realizing ${\bf Z}_3$ outer automorphism symmetry which 
is not present in 10 dimensions. The implication of the 
above discussion is that one needs relatively simple rules to 
facilitate model building. Such rules have been
 derived \cite{kt} within the framework of 
{\em asymmetric orbifolds} \cite{vafa}.

{}With the appropriate model building tools available, it became possible 
to construct \cite{kt,kt10} and classify \cite{class5,review} 
3-family grand unified string models within the framework of 
asymmetric orbifolds in perturbative heterotic string theory. 
Here we briefly discuss the results of this classification. 
For each model we list here there are 
additional models connected to it via classically flat 
directions \cite{class5}:\\
$\bullet$ One $E_6$ model with 5 left-handed and 2 
right-handed families, and asymptotically free $SU(2)$ hidden sector 
with 1 ``flavor''.\\ 
$\bullet$ One $SO(10)$ model with 4 left-handed and 1 right-handed 
families, and $SU(2)\otimes SU(2)\otimes SU(2)$ hidden sector which 
is {\em not} asymptotically free at the string scale.\\
$\bullet$ Three $SU(6)$ models:\\
({\em i}) The first model has 6 left-handed and 3 right-handed families, 
and asymptotically free $SU(3)$ hidden sector with 3 ``flavors''.\\
({\em ii}) The second model has 3 left-handed and no right-handed families, and asymptotically free $SU(2)\otimes SU(2)$ hidden sector with matter content consisting of doublets of each $SU(2)$ subgroup as well as bi-fundamentals.\\
({\em iii}) The third model has 3 left-handed and no right-handed families, and asymptotically free $SU(4)$ hidden sector with 3 ``flavors''. As far as the observable sector is concerned, this model is a minimal $SU(6)$ extension of the minimal $SU(5)$ model \cite{Georgi}.\\
$\bullet$ Finally, there are some additional $SU(5)$ models which do not seem to be phenomenologically appealing (see below).

{}All of the above models share some common phenomenological features. Thus, there is only one adjoint and no other higher dimensional Higgs fields in all of these models. The $E_6$ and $SO(10)$ models 
(and other related models) do {\em not} possess anomalous $U(1)$. All three $SU(6)$ models listed above {\em do} have anomalous $U(1)$  (which in string theory is broken via the Green-Schwarz mechanism \cite{GSDSW}). The above models all possess non-Abelian hidden sector. There, however, exist models
where the hidden sector is completely broken.

{}To study phenomenological properties of these models it is first necessary to deduce tree-level superpotentials for them. This turns out to be a rather non-trivial task as it involves understanding scattering in asymmetric orbifolds. There, however, are certain simplifying circumstances here due to the fact that asymmetric orbifold models possess enhanced discrete and continuous gauge symmetries. Making use of these symmetries,
 the tools for computing tree-level superpotentials for a class of asymmetric orbifold models (which includes the models of interest for us 
here) have been developed in \cite{kst}. The perturbative 
superpotentials for some of the three-family grand unified 
string models were computed in \cite{kst,review}.

{}The knowledge of tree-level superpotentials allows one to analyze 
certain phenomenological issues such as proton stability (doublet-
triplet splitting and
$R$-parity violating terms) and Yukawa mass matrices. The question 
of supersymmetry breaking can also be addressed \cite{review} by augmenting the 
tree-level superpotentials with non-perturbative contributions which 
are under control in $N=1$ supersymmetric field theories \cite{Seiberg}.  

{}Thus, in \cite{kstv} doublet-triplet splitting problem and Yukawa
mass matrices were studied for the $SO(10)$ models.
It was found that certain degree of
fine-tuning is required to solve the doublet-triplet splitting
problem, suppress dangerous $R$-parity violating terms and achieve
realistic Yukawa mass matrices.
All $SU(5)$ models suffer from severe fine-tuning problem stemming from
 the doublet-triplet splitting as there are no ``exotic'' higher
dimensional Higgs fields among the light degrees of freedom.
The latter are required by all known field theory solutions to the
problem.

{}Similar issues for the $SU(6)$ models were studied in \cite{review}. The 
results of \cite{review} indicate that the doublet-triplet 
splitting does not seem to be as big of a problem for the $SU(6)$ 
models as it is for their $SO(10)$ and $SU(5)$ counterparts. 
Thus, the doublet-triplet splitting in the $SU(6)$ models might be solved via the pseudo-Goldstone mechanism developed in the field theory context \cite{pseudo}.
However, the troubles 
with $R$-parity violating terms and Yukawa mass matrices still persist for 
these models.

{}In \cite{review} the possible patterns of 
supersymmetry breaking 
in the three-family grand unified string models were also analyze. 
It was found that the supersymmetry breaking scale in these 
models comes out either too high to explain the electroweak 
hierarchy, or below the 
electroweak scale unless some degree of fine-tuning is involved.

{}Since none of the three-family grand unified string models 
constructed to date appear to be phenomenologically flawless, 
one naturally wonders whether there may exist (even within 
perturbative heterotic vacua) other such models 
with improved phenomenological characteristics.
Thus, all {\em a priori} 
possible free-field embeddings of higher-level current algebras 
within perturbative heterotic superstring framework have been 
classified \cite{dien}. This, however, does not guarantee that 
any given embedding can be incorporated in a consistent string model,
 and even if this is indeed possible, there need not exist 
three-family models within such an embedding. The three-family 
grand unified string models discussed in this review are 
concrete realizations of the {\em diagonal} level-3 embedding 
$G_{diag}\supset G\otimes G\otimes G$. However, even if our 
models do exhaust all three-family grand unified string models 
within free-field realized perturbative heterotic superstring, 
there may exist non-free-field 
grand unified string models with three families. Tools for 
constructing such models are way underdeveloped at present, so that 
for years to come the asymmetric orbifold models we discuss in this 
review might be the only ones available. Regardless of their 
phenomenological viability they provide the {\em proof of existence} 
for three-family grand unified string models, and can serve as 
a stringy paradigm for such model building in general, and also 
give insight to the ``bottom-up'' approach.   

{} This work was supported in part by the grant NSF PHY-96-02074, and the DOE 1994 OJI award. We would also like to thank Albert and Ribena Yu for financial support.

\end{document}